\documentclass[onecolumn,showpacs,showkeys,preprintnumbers,floatfix,aps]{revtex4}

\usepackage{palatino}

\usepackage{amsmath}
\usepackage{amssymb}

\usepackage{graphicx}
\usepackage{epsfig}
\usepackage{rotating}

\usepackage{dcolumn}
\usepackage{bm}

\begin{document}


\title{%
$S$--pairing in neutron matter \\
I. Correlated Basis Function Theory
}%

\author{Adelchi~Fabrocini}
\email{Adelchi.Fabrocini@df.unipi.it}
\affiliation{%
   Dipartimento di Fisica ``Enrico Fermi'', Universit\`a di Pisa
}%
\affiliation{%
   INFN, Sezione di Pisa, I-56100 Pisa, Italy
}%

\author{Stefano~Fantoni}
\email{fantoni@sissa.it}
\affiliation{%
   International School for Advanced Studies, SISSA
}%
\affiliation{%
   INFM DEMOCRITOS National Simulation Center,
   I-34014 Trieste, Italy
}%

\author{Alexey~Yu.~Illarionov}
\email{illario@sissa.it}
\affiliation{%
  International School for Advanced Studies, SISSA,
	I-34014 Trieste, Italy
}%
\affiliation{%
  INFN, Sezione di Pisa, I-56100 Pisa, Italy
}%

\author{Kevin~E.~Schmidt}
\email{Kevin.Schmidt@asu.edu}
\affiliation{%
   Department of Physics and Astronomy,
   Arizona State University, Tempe, AZ, 85287
}%
\affiliation{%
   International School for Advanced Studies, SISSA,
   I-34014 Trieste, Italy
}%

\date{\today}

\begin{abstract}
$S$--wave pairing in neutron matter is studied within an extension of
correlated basis function (CBF) theory to include the strong, short range
spatial correlations due to realistic nuclear forces and the
pairing correlations of the Bardeen, Cooper and Schrieffer (BCS)
approach. The correlation operator
contains central as well as tensor components. The correlated
BCS scheme of Ref.~\cite{Fantoni:1981}, developed for simple scalar
correlations, is generalized to this more realistic case. The energy
of the correlated pair condensed phase of neutron matter is
evaluated at the two--body order of the cluster expansion, but considering
the one--body density and the corresponding energy vertex corrections
at the first order of the Power Series expansion. Based on these
approximations, we have derived a system of Euler equations for the
correlation factors and for the BCS amplitudes, resulting in correlated
non linear gap equations, formally close to the standard
BCS ones. These equations have been solved for the momentum independent
part of several realistic potentials (Reid, Argonne $v_{14}$ and
Argonne $v_{8^\prime}$) to stress the role of the tensor correlations and of
the many--body effects. Simple Jastrow
correlations and/or the lack of the density corrections enhance the gap
with respect to uncorrelated BCS, whereas it is reduced
according to the strength of the tensor interaction and following the
inclusion of many--body contributions.
\end{abstract}



\maketitle

\section{Introduction}
\label{sec:Intro}

Superfluidity in neutron matter
has been a fascinating topic in many--body physics and astrophysics ever since
Migdal~\cite{Migdal:1960} proposed the possibility of superfluid matter
in neutron stars.
In the inner crust of the star, $^1S_0$
pairing in the low density neutron gas permeating the lattice of
neutron rich nuclei may occur and peak at densities much lower than 
the empirical nuclear matter saturation density, $\rho_0 =
0.16~\text{fm}^{-3}$.
A similar pairing may take place for the low concentration proton component
in the highly asymmetrical nuclear matter in the star's interior.
At higher interior densities, neutrons may also pair in the anisotropic
$^3P_2$--$^3F_2$ partial wave. A realistic evaluation of the density
regimes
where superfluidity takes place and of the strength of the connected energy
gaps is needed for a quantitative understanding of important features
of neutron stars, such as the cooling rate\cite{Tsuruta:1998,Heiselberg:2000}
and the post--glitch relaxation times~\cite{Sauls:1989,Pines:1992}.

The qualitative aspects of superfluidity were shown to be describable
in nuclei~\cite{Bohr:1958} and in infinite systems of interacting
fermions~\cite{Cooper:1959} by the extension of the
theory of superconductivity of Bardeen, Cooper and
Schrieffer~\cite{Bardeen:1957} (BCS). In terms of the nucleon--nucleon
(NN) interaction, it is the long range attraction of the $^1S_0$ NN potential
that dominates in the inner crust density regime,
allowing for $S$--wave pairing. The gap closes with rising density since
the short range repulsion is more and more effective. Proton superfluidity
(or superconductivity) has a similar origin in the interior, while
higher density $^3P_2$--$^3F_2$ neutron pairing is traced back to
non central, tensor and spin--orbit, components. In BCS theory
Cooper--like pairs allow for superfluidity even in presence of the
short range repulsion of modern potentials.

On the other hand, the strong nuclear interaction induces short range
correlations in the wave function, which also largely screen the
repulsion and introduce many--body contributions. These two features
have competing effects, since the former is expected to increase the gap,
whereas the latter may diminish it. Modern many--body theories, such as
the method of correlated basis functions~\cite{Feenberg} (CBF), the
Bethe--Brueckner--Goldstone expansion~\cite{BBG} (BBG), the
self--consistent
Green's functions theory~\cite{SCGF} (SCGF), and lately quantum Monte
Carlo~\cite{QMC} (QMC), can, with efficiency and accuracy, deal with short
range correlations in
normal phase nucleonic matter. It can be reasonably expected that these
methods also may be able to provide a similarly realistic description
of the superfluid phase, especially when modern NN potentials are used.

Within CBF the short range correlations
are introduced by acting with a many--body correlation
operator on a set of model functions, so defining a correlated basis to be used
in a perturbative expansion where the highly non perturbative short range
correlation effects are already embedded in the basis. The zeroth order
of the correlated perturbative expansion corresponds to a variational
approach, since the correlation operator (and the ground-state
model wave function)
can be derived applying the Ritz variational principle. The variational
level may already give reliable results if the correlation operator is
chosen in an appropriate way.
Because realistic NN potentials have important spin-- and isospin--dependent
components, both central and non central (e.g the tensor potential,
mainly originating from one--pion exchange), a good variational choice for
the pair correlation $\widehat{f}(ij)$ must include at least six
components, 
\begin{equation}
\widehat{f}_6(ij) \ = \ \sum_{p=1,6} f^{(p)}(r_{ij})
\widehat{O}^{(p)}(ij) \ ,
\label{eq:f6}
\end{equation}
where
 $\widehat{O}^{(p = 1,2,3)}(ij) = 1$,
 $\mathbf{\sigma}(i) \cdot \mathbf{\sigma}(j)$,
 $\widehat{S}(ij) = (3 \hat r_\alpha(ij) \hat r_\beta(ij) - \delta_{\alpha
\beta})
          \sigma_\alpha(i)\sigma_\beta(j)$,
and
 $\widehat{O}^{(p^\prime = p+3)}(ij) =
  \widehat{O}^{(p)}(ij) \otimes \mathbf{\tau}(i) \cdot \mathbf{\tau}(j)$.
The greek indices denote the Cartesian components.
This choice of the operatorial
dependence of the correlation is consistent with the use of the
non central and momentum independent $v_6$ potentials of the form,
\begin{equation}
	\widehat{v}_6(ij) \ = \ \sum_{p=1,6} v^{(p)}(r_{ij})\widehat{O}^{(p)}(ij) \ .
\label{eq:v6}
\end{equation}
However, $\widehat{f}_6(ij)$ is in general a very good variational choice
\emph{for all} the realistic potentials.
 The introduction of such structures directly in the correlation operators
allows the variational approach to describe microscopically the
structure of nuclear matter~\cite{WFF} and finite
nuclei~\cite{CBF_in_nuclei}
with a good accuracy.

In this paper we are only dealing with pure neutron matter (PNM), therefore
$\mathbf{\tau}(i) \cdot \mathbf{\tau}(j) \equiv 1$
and the 6--operator algebra underlying
$\widehat{v}_6(ij)$
and
$\widehat{f}_6(ij)$
reduces to the first 3 components
$p = 1,2,3$,
where
$f_\text{PNM}^{(p)} = f^{(p)} + f^{(p + 3)}$
and
$v_\text{PNM}^{(p)} = v^{(p)} + v^{(p + 3)}$.

Since the operators in (\ref{eq:f6}) do not commute, 
the many--body correlation operator, $\widehat{F}_6(1,2,..N)$, is given by
the symmetrized product, 
\begin{equation}
\widehat{F}_6(1,2,..N) \ = \
  \mathcal{S} \left[ \prod_{i < j = 1,N} \widehat{f}_6(ij) \right] \ .
\label{eq:F6}
\end{equation}
In CBF theory such operators are kept fixed for all the intermediate states.
The correlated CBF intermediate states are obtained by acting with 
$\widehat{F}_6(1,2,..N)$ on the
corresponding uncorrelated Slater determinant.

 An alternative approach, hereafter denoted as CBF-J, consists in starting
with a simpler Jastrow correlation~\cite{Feenberg}, depending only on the 
interparticle distance,
\begin{equation}
F_J(1,2,..N) \ = \ \prod_{i<j=1,N} f_J(r_{ij}) \ ,
\label{eq:jastrow}
\end{equation}
and introducing the spin/isospin dependence via a Jastrow--correlated
perturbative expansion~\cite{Chen:1986}. 
This choice may not be very efficient since the whole spin-isospin dependence 
must be perturbatively included. However, 
the terms of the CBF-J expansion have a much simpler structure than
those of the CBF expansion, based on spin dependent correlation operators, and can be computed by Fermi hypernetted chain 
(FHNC) resummation \cite{Fantoni:1981}.
A possible drawback of the CBF-J perturbative expansion is the complexity of going beyond the second order perturbation level
which may be insufficient in the Jastrow-like CBF theory.

Variational CBF theory has been applied to the $S$--wave nucleonic
superfluid in Ref.~\cite{Chen:1993} using central potentials and
correlations, without tensor components,
\begin{eqnarray}
\widehat{v}_4(ij) \ = \ \sum_{S,T=0,1} v^{(ST)}(r_{ij})
\widehat{P}^{(ST)}(ij) \ ,
\label{eq:v4} \\
\widehat{f}_4(ij) \ = \ \sum_{S,T=0,1} f^{(ST)}(r_{ij})
\widehat{P}^{(ST)}(ij) \ ,
\label{eq:f4}
\end{eqnarray}
where $\widehat{P}^{(ST)}(ij)$ are projectors onto the two--body subspace of
total spin--isospin $ST$. The $N$--body correlation
operator is then given by:
\begin{equation}
\widehat{F}_4(1,2,..N) \ = \
 \mathcal{S} \left[ \prod_{i<j=1,N} \widehat{f}_4(ij) \right] \ .
\label{eq:F4}
\end{equation}
Lowest order cluster expansion was used to derive a correlated
gap equation with the $\widehat{v}_4$ version of the Reid soft core NN
interaction~\cite{Reid:1968,Pandharipande:1979}.
This correlated theory was developed within the independent Cooper
pairs approximation and does not consider the dependence of the
correlation on the BCS amplitudes. The approach
takes essentially into account the screening of the core repulsion due to the 
repulsive part of the correlation,
and leads to a larger gap than BCS.
Chen et al.~\cite{Chen:1986} studied $S$--pairing with the
Reid ${v}_6$ potential, including the interaction tensor components,
using the independent Cooper pairs approximation.
They considered a simple Jastrow correlation rather then the correlation operator
$\widehat{F}_6$ of Eq.~\ref{eq:f6},
but computed the variational energy at
a higher level of the cluster expansion through 
FHNC theory~\cite{FHNC} . A reduction of the BCS gap of about
$30\%$ was found, attributable, however, to a rather poor choice of
the Jastrow factor. The authors of Ref.~\cite{Chen:1986}
also computed the second order perturbative CBF correction to the
pairing matrix element on top of the Jastrow estimate.
This approach, which should take into account \emph{medium polarization},
led to a dramatic reduction of the gap by
$\sim 80\%$, much larger than all the other estimates of the polarization
effects, and inconsistent with X--ray
observations~\cite{Alpar:1987}. Inspite of the fact that the matrix
elements of CBF perturbation theory are easier to compute in a Jastrow correlated basis,
its convergence for large non--central potentials in such a basis is still to be assessed.

The independent Cooper pairs approximation was overcome in ref.~\cite{Fantoni:1981}, hereafter denoted as I,
with a Jastrow fully correlated BCS theory.
In this work we begin to extend the work of I to the case of correlations having spin--isospin dependent,
with both central and tensor components ($f_6$ model).

The use of a $f_6$ correlation does not allow for a complete sum of the FHNC diagrams, very much the same as for the case of normal phase.
Similarly to that case the massive resummations of diagrams can be performed
using the single operator chain (SOC) approximation of Ref.~\cite{Pandharipande:1979}.
In this paper we limit our attention to study pure neutron matter at the two--body level plus vertex corrections of the cluster expansion of
$\langle \hat{H} - \mu \hat{N} \rangle$,
where $\mu$ is the chemical potential determined by fixing the correct mean value of the particle number operator (or the density, for infinite systems)
 $\langle \hat{N} \rangle = \sum_m \langle a^{\dagger}_m a_m \rangle$.

The one--body density
$\rho = \langle\hat N\rangle/\Omega$,
and consequently the vertex corrections in
$\langle \hat{H} \rangle$, 
will be here computed at the first order of the Power Series expansion~\cite{FHNC}.
This approximation guarantees in the normal phase the correct density normalization, order by order, and introduces a first flavor of many--body effects.
The expectation value
$\langle \hat{H} \rangle$
will be computed at the second order of the cluster expansion, which provides a sufficiently good description of the short--range correlations.

Minimization of
$\langle \hat{H} - \mu \hat{N} \rangle_2$
with respect to the correlation functions
$f_6$
and to the BCS amplitudes leads to a coupled set of Euler and gap equations, which we denote as correlated BCS equation.
The solution of such equation is a preliminary, very important step towards a full calculation,
which will include higher order effects in the evalution of both
$\langle \hat{N} \rangle$
and
$\langle \hat{H} \rangle$
and second order perturbative corrections following orthogonal CBF theory of ref.~\cite{CBF}.
A second approach consists in using the Auxiliarly Field Diffusion Monte Carlo (AFDMC) method
to calculate the gap energy of a finite number of neutrons in the superfluid phase.
Such a method has been used to simulate up to $114$ neutrons in a periodical box
to evaluate the equation of state at zero temperature in neutron matter in the normal phase~\cite{AFDMC_nm}.
The extension to superfluid phases can be done using the method developed in the recent work~\cite{Carlson:2003}
in the study of low density Fermi gas in the regime of large scattering length interaction.
AFDMC simulations of this type crucially depend upon the choice of a guiding function to fix the nodes and the phases of the wave function.
Therefore, the BCS amplitudes resulting from solving the correlated BCS equation are a fundamental input to the AFDMC simulations.
Preliminary results of that simulations performed with $14$ neutrons have already been published~\cite{PRL_BCS}.
Besides the derivation of the correlated BCS equation and its solution for several potentials of the
$v_6$
type (like the truncated versions of the Reid~\cite{Reid:1968}, Argonne
$v_{14}$
and Argonne
$v_8$~\cite{v14}
potentials) we have evaluated the gap energy with and without vertex corrections.
The latter to compare with BBG~\cite{BBG} and SCGF~\cite{SCGF},
the former to estimate the effect of the three--body terms of which the vertex corrections are the main part.
%

The plan of the paper is as follows: in Section 2 the correlated BCS theory
for a $f_6$ correlation is presented; Section 3 contains the Euler and
correlated gap equations; numerical results and details on the solution of
the equations are given in Section 4; Section 5 will briefly discuss our
results and give conclusions and perspectives

\section{Correlated BCS theory}
\label{sec:Corr_BCS}

 A correlated wave function for the neutron matter superfluid phase
is constructed as
\begin{equation}
|\Psi_s\rangle \ = \ \hat F \ |\text{BCS}\rangle \ ,
\label{eq:psi_s}
\end{equation}
where the model BCS--state vector is
\begin{equation}
|\text{BCS}\rangle \ = \ \prod_{\bm k}
( u_{\bm k} + v_{\bm k}a^\dagger_{{\bm k}\uparrow}
a^\dagger_{-{\bm k}\downarrow})|0\rangle \ .
\label{eq:BCS}
\end{equation}
$u_{\bm k}$ and $v_{\bm k}$ are the real variational BCS amplitudes,
satisfying the relation $u^2_{\bm k} + v^2_{\bm k} = 1$,
$|0\rangle$ is the vacuum state and $a^\dagger_m$ is the fermion
creation operator in the single--particle state
$|m = \bm{k}, \sigma \rangle$
whose wave function is
\begin{equation}
	\langle x \equiv \bm{r}, s | a_m^\dagger | 0 \rangle =
	\phi_{m\equiv {\bm k}, \sigma}(x\equiv {\bm r},s) \ = \
	\dfrac{1}{\sqrt{\Omega}} \ \eta_\sigma (s) \exp(\imath {\bm k}\cdot {\bm r}) \ .
\label{eq:single}
\end{equation}
$\Omega$ is the normalization volume and
$\eta_{\sigma=\uparrow,\downarrow}(s)$ is the spin wave function with
spin projection $\sigma$. The second--quantized correlation operator
$\hat F$ is written in terms of the N--particle correlation operators,
$\hat F_N$, as
\begin{equation}
\hat F \ = \ \sum_{N, m_N} \ \hat F_N \
|\Phi_N^{m_N}\rangle\langle\Phi_N^{m_N}| \ ,
\label{eq:hat_F}
\end{equation}
where
$m_N$
specifies a set of $N$ single--particle states with
$N = 0,2,4,\dots$
In coordinate representation and for a $f_6$--type correlation we have:
\begin{equation}
\langle x_1,x_2,..x_N|\hat F_N | \Phi_N^{m_N}\rangle\ \ = \
\widehat{F}_6(1,2,..N)
\left\{ \phi_{m_1}(x_1)\phi_{m_2}(x_2)..\phi_{m_N}(x_N)\right\}_A \ .
\label{eq:F_N}
\end{equation}
The suffix $A$ stands for an antisymmetrized product of single--particle
wave functions and $\widehat{F}_6(1,2,..N)$ is the $f_6$ N--particle 
correlation operator~(\ref{eq:F6}). 

In I the cluster expansions of the two--body distribution function,
$g(r_{12})$,
\begin{equation}
g(r_{12}) \ = \ \dfrac{1}{{\mathcal N}\rho^2} \
\sum_{\sigma_1,\sigma_2} \
\langle \Psi_s|
\Psi^\dagger_{\sigma_1}({\bm r}_1)\Psi_{\sigma_1}({\bm r}_1)
\Psi^\dagger_{\sigma_2}({\bm r}_2)\Psi_{\sigma_2}({\bm r}_2)
|\Psi_s \rangle \ ,
\label{eq:g2}
\end{equation}
and of the one--body density matrix,
$n(r_{11^\prime})$,
\begin{equation}
n(r_{11^\prime}) \ = \ \dfrac{1}{\mathcal N} \,
\sum_{\sigma_1,\sigma_{1^\prime}} \
\langle \Psi_s|
\Psi^\dagger_{\sigma_1}({\bm r}_1)\Psi_{\sigma_{1^\prime}}({\bm
r}_{1^\prime})
|\Psi_s \rangle \ ,
\label{eq:OBDM}
\end{equation}
in the Jastrow correlated case were studied.
In the above equations, $\cal N$ are normalization constants,
$\Psi_{\sigma}({\bm r})$
and
$\Psi^{\dagger}_{\sigma}({\bm r})$
are the destruction and creation field operators.

In I it was proved that
$g(r_{12})$ and $n(r_{11^\prime})$ are given by the sum of all the linked
cluster diagrams, constructed by the dynamical correlation lines
($h_J = f_J^2 - 1$ for the Jastrow correlation) and the BCS
statistical correlations,
\begin{align}
l_v(r) =& \dfrac{\nu}{\rho_0} \int \dfrac{d^3k}{(2\pi)^3}
\ \exp ({\imath {\bm k} \cdot {\bm r}})\ v^2(k) \ ,
\label{def:l_v} \\
l_u(r) =& \dfrac{\nu}{\rho_0} \int \dfrac{d^3k}{(2\pi)^3}
\ \exp ({\imath {\bm k} \cdot {\bm r}})\ u(k) v(k) \ ,
\label{def:l_u}
\end{align}
where
$\nu$ is the spin--isospin degeneracy ($\nu = 2$ for PNM) and
$\rho_0$ is the average density of the uncorrelated BCS model, given by
\begin{equation}
\rho_0 \ = \ \nu \int \dfrac{d^3k}{(2\pi)^3} v^2(k) \ .
\label{eq:rho0}
\end{equation}
The FHNC equations, derived in I, sum at all orders the cluster diagrams
contributing to $g(r_{12})$ and $n(r_{11^\prime})$ in the Jastrow case.
Here we are dealing with a spin--dependent correlation operator of the type
$\widehat F_6$
of Eq.~(\ref{eq:F6}), reduced to the PNM case.

In addition to the complexity introduced by the spin--dependence, the noncommutativity of
$\hat f(ij)$
with
$\hat f(ik)$
implies that any given cluster diagram generates as many clusters as the number of possible ordering of the operators presented in the diagram.
This is a formidable task, which is not been yet solved.
Reasonable approximations have been devised \cite{Pandharipande:1979,Akmal:1998} to sum up the leading cluster terms.
Instead of following such schemes we calculate exactly the lowest order correlated cluster terms of
$g(r_{12})$
and
$n(r_{12} = 0)$.
This is justified be the fact that we consider short--range correlations and a low density system.
Moreover, we are mainly interested to derive the correlated BCS equations.

It is well known that normalization properties are better approximated by the succesive terms of the power series expansion \cite{Fantoni_lecture:1998}
namely the expansion in the number of correlation lines.
The energy expectation value is instead better evaluated using the expansion in the number of particles, or, equivalently, in the density.
Such inconsistency can be partionaly resolved by performing a full FHNC summation of both quantities
in the case if the elementary diagrams give the negligible contributions.
Here we will calculate
$n(r_{12} = 0)$
up to the first orger of PS expansion and
$g(r_{12})$
at the two--body cluster level
plus the vertex corrections, evaluated at the first order of the PS expansion, to be consistent with
$n(r_{12} = 0)$.


\subsection{One--body density and vertex corrections}
\label{sub:OBD}

For a BCS--type trial function the density is given by:
\begin{equation}
\rho = \dfrac{\langle \hat{N}\rangle }{\Omega} \ =
\dfrac{\sum_m <a^{\dagger}_m a_m>}{\Omega} \ .
\label{eq:OBD}
\end{equation}

Fluctuations with respect to this average vanish in the thermodynamic limit.
We stress that the actual density,
$\rho$,
differs from
$\rho_0$
because the correlation operator affects
$\langle \hat{N}\rangle $
(see I). Therefore,
$\rho_0$
has to be considered as a \emph{variational} parameter, and
$\rho$
has to be computed self--consistently.

\begin{figure}[h]
	\includegraphics[angle=0, scale=0.6]{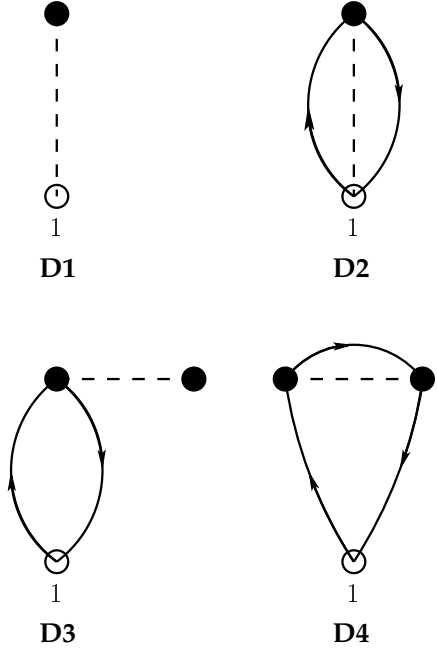}
 \caption{\label{Pic:BCS_diagrams}%
The lowest order diagrams that contribute to the vertex corrections.
}%
\end{figure}

The calculation of
$\rho$
follows the FHNC scheme of I.
We limit our attention to the FHNC diagrams with zero and one correlation lines,
e.g. those belonging to the first order of the Power Series cluster expansion,
Fig.~(\ref{Pic:BCS_diagrams}) shows the first order diagrams.

The external point, denoted as $1$, is represented by an open dot, whereas
the internal points are given as black dots.
The oriented lines represent exchange
$l_u$
or
$l_v$
functions, composed as in I, whereis the dashed ones are dynamical correlations
$\widehat F^2 - 1$.
Diagram $D4$ has an overall $1/2$ symmetry factor, canceling the $2$ factor
coming from the two exchange loops with oposite orientations.

In the standard FHNC theory for the normal phase,
diagrams $D1$--$D4$ add up to give zero contribution: $D1$ is canceled by $D3$ and $D2$ by $D4$.
We are left with the uncorrelated zeroth order diagrams, given rise to the Fermi gas momentum distribution,
$n_k = \Theta(k_\text{F} - k)$,
$k_\text{F} = (6\pi^2\rho /\nu)^{1/3}$
being the Fermi momentum.
The total density correctly coincides, at any order of the power series, with that of the uncorrelated Fermi sea,
$\rho=\rho_0$.

In correlated BCS theory this cancellation no longer holds, and corrections to the uncorrelated
$\rho_0$
are found, namely
$\rho \neq \rho_0$.

 
Following the notation of I, diagram $D1$ is the first order of the
vertex correction of the $U_d$ type, and the other three diagrams 
are included into the vertex correction of the $U_e$ type.
Keeping only the linear terms, the density $\rho$ is given by
\begin{equation}
\rho \sim \rho_1 \ = \ c_d \rho_0 \ ,
\label{eq:rho1}
\end{equation}
where $c_d$ is given, in terms of the cluster terms $U_d$ and $U_e$,
 by the following equations:
\begin{align}
 c_e =& \ 1 + U_d \ ,
\nonumber \\
 c_d =& \ c_e + U_e = 1 + U_d + U_e \ .
\end{align}
Accordingly, $c_e$ is the correction to vertices which are arrival points
of some exchange lines, whereas the complete $c_d$ is
the correction for vertices only connected to dynamical lines.

It is straightforward to extend the algebraic methods given in I to the case of the spin--dependent correlations.

We will now discuss the terms associated with the diagrammatic structures ($D1$--$D4$), contributing to
$U_d$
and
$U_e$.
For $D1$ we get:
\begin{equation}
D1 \rightarrow U_d = \rho_0 \int d^3r \sum_{i,j} K^{ij1}
 \left(f^{(i)}(r)f^{(j)}(r) - \delta_{i1}\delta_{j1}\right) \ .
\label{eq:Ud}
\end{equation}
The $K^{ijk}$ matrix is given by
\begin{gather}
	K^{ij1} = \left(
	\begin{array}{ccc}
		1 & 0 & 0 \\
		0 & 3 & 0\\
		0 & 0 & 6 \\
	\end{array}
	\right) \ , \
	K^{ij2} = \left(
	\begin{array}{ccc}
		0 & 1 & 0 \\
		1 & -2 & 0\\
		0 & 0 & 2 \\
	\end{array}
	\right) \ , \
	K^{ij3} = \left(
	\begin{array}{ccc}
		0 & 0 & 1 \\
		0 & 0 & 1 \\
		1 & 1 & -2 \\
	\end{array}
	\right) \ .
	\label{Kijk}
\end{gather}

Two terms are associated to $D3$: the first term has both $l_v$-type
exchange lines; in the second term one line is of the $l_u$-type.
The total contribution is:
\begin{equation}
D3 \rightarrow U_{e3} = -U_d \dfrac{\nu}{\rho_0} \int
\dfrac{d^3k}{(2\pi)^3}
 \left(v^4(k) - u^2(k) v^2(k) \right) \ .
\end{equation}
In the normal phase, $u(k) v(k) = 0$ and
$v^2(k) = \Theta(k_\text{F} - k)$, and $U_d + U_{e3} = 0$.

Similarly to $D3$, $D2$ has two analogous terms:
\begin{equation}
D2 \rightarrow U_{e2} = U_{e2}^v + U_{e2}^u \ ,
\end{equation}
where
\begin{align}
U_{e2}^v =& -\dfrac{\rho_0}{\nu} \int d^3r \ l^2_v(r) \sum_{i,j,k}
 K^{ijk} A^k P_k \left( f^{(i)}(r)f^{(j)}(r)-\delta_{i1}\delta_{j1} \right)
\ , \\
U_{e2}^u =& \dfrac{\rho_0}{\nu} \int d^3r \ l^2_u(r) \sum_{i,j,k}
 K^{ijk} A^k B_k \left( f^{(i)}(r)f^{(j)}(r)-\delta_{i1}\delta_{j1}
\right))
\ ,
\end{align}
with $A^k = (1,3,6)$, $P_k = (1,1,0)$ and $B_k = (1,-1,0)$
(note that the factor $1/2$ of the spin--exchange operator is
included in the factor in front of the integrals).

Diagram D4 has three different exchange patterns giving the contributions:
\begin{itemize}
\item[(i)]
$U_{e4}^{vv}$, having all $l_v$--type exchanges,
\begin{equation}
U_{e4}^{vv} = \dfrac{\rho_0}{\nu} \int d^3r \ l_v(r) \, l_{vv}(r) \
 \sum_{i,j,k}
 K^{ijk} A^k P_k \left( f^{(i)}(r)f^{(j)}(r)-\delta_{i1}\delta_{j1}
\right) \ ,
\end{equation}
with,
\begin{equation}
l_{vv}(r) = \dfrac{\nu}{\rho_0} \int \dfrac{d^3k}{(2\pi)^3}
 \exp (\imath {\bm k} \cdot {\bm r}) v^4(k) \ .
\end{equation}
Again, if $v^2(k) = \theta(k_\text{F} - k)$, then $U_{e2}^v +
U_{e4}^{vv} = 0$.

\item[(ii)]
$U_{e4}^{uu}$, having two $l_u$ exchanges joining at the external point
$1$,
 while the third exchange line is of the $l_v$--type,
\begin{equation}
U_{e4}^{uu} \ = \ - U_{e2}^{v} - U_{e4}^{vv}
\end{equation}

\item[(iii)]
$U_{e4}^{uv}$, having a $l_u$ exchange joining with a $l_v$ one at point
$1$,
the third line being of the $l_u$--type,
\begin{equation}
U_{e4}^{uv} = \dfrac{\rho_0}{\nu} \int d^3r \ l_u(r) \, l_{uv}(r) \
 \sum_{i,j,k}
 K^{ijk} A^k B_k \left( f^{(i)}(r)f^{(j)}(r)-\delta_{i1}\delta_{j1}
\right) \ ,
\end{equation}
with,
\begin{equation}
l_{uv}(r) = \dfrac{\nu}{\rho_0} \int \dfrac{d^3k}{(2\pi)^3}
 \exp (\imath {\bm k} \cdot {\bm r}) u(k) v^3(k) \ .
\end{equation}
\end{itemize}

The total $U_{e4}$ term is the sum
\begin{equation}
U_{e4} = U_{e4}^{vv} + U_{e4}^{uu} + 2 U_{e4}^{uv} \ .
\label{eq:Ue4}
\end{equation}

In conclusion, $U_e$ is given by:
\begin{equation}
U_e = U_{e2} + U_{e3} + U_{e4} = U_{e2}^u + U_{e3} + 2 U_{e4}^{uv} \ .
\label{eq:Ue}
\end{equation}

\subsection{Potential energy}
\label{sub:PotEn}

We perform the calculation of the expectation value of a
$v_6$
potential at the two-body order for the cluster expansion,
but including also the vertex corrections at the interaction points,
$1$
and
$2$.
The reason for going beyond the simple two--body approximation in
the superfluid phase lies in the correlation driven modification of
the expectation value of the number operator with respect to BCS,
as discussed in the previous subsection.

The vertex corrections lead to a fully factorized term, similiar to that in the Jastrow correlated BCS case of I,
plus commutator correction terms,
 
\begin{align}
\dfrac{\langle\hat{V}\rangle_2}{\langle\hat{N}\rangle_1} = V_2 =&
 \dfrac{\rho}{2} \int d^3r_{12} \left[ V_d(r_{12}) -
 \dfrac{1}{\nu} \left(\dfrac{c_e}{c_d}\right)^2
 \left[ V_{ev}(r_{12}) l_v^2(r_{12}) -
 V_{eu}(r_{12}) l_u^2(r_{12}) \right] \right]
\nonumber \\
 +& \Delta C_d + \Delta C_{ev} + \Delta C_{eu} \ ,
\label{eq:pot}
\end{align}
where the $\Delta C$ terms are the commutator corrections.
 
The $V_d$ and $V_{ev}$ functions coincide with the direct
and exchange terms of the normal phase of PW,
\begin{align}
V_d(r) \ =& \
 \sum_{i,j,k} f^{(i)}(r) v^{(j)}(r) f^{(k)}(r) K^{ijk} A^k \ ,
\nonumber \\
V_{ev}(r) \ =& \
  \sum_{i,j,j^\prime,k,l} f^{(i)}(r) v^{(j)}(r) f^{(k)}(r)
 K^{ijj^\prime} K^{j^\prime kl} A^l P_l \ .
\label{eq:vnormal}
\end{align}

After performing the spin algebra corresponding to the last term of
the first line of (\ref{eq:pot}), we obtain

\begin{equation}
V_{eu}(r)\ = \
 \sum_{i,j,j^\prime,k,l} f^{(i)}(r) v^{(j)}(r) f^{(k)}(r)
 K^{ijj^\prime} K^{j^\prime kl} A^l B_l \ .
\label{eq:ucomponent}
\end{equation}

All the vertex structures $D1$--$D4$ contribute to
$c_d$, as discussed in the previous subsection. However, only
$D1$ and $D2$ originate commutator contributions.

The commutator terms
$\Delta C$
are calculated following the algebraic methods of ref.~\cite{Pandharipande:1979}.
After some lengthy calculations we obtain:
\begin{align}
\Delta C_d &=
- (\bar{U}_d - \bar{U}_{e2}^v + \bar{U}_{e2}^u) \rho_0 \int d^3r_{12}
 \nonumber \\
&\times
\sum_{i,j,k} K^{ijk}A^k (3-\delta_{i1}-\delta_{j1}-\delta_{k1})
f^{(i)}(r_{12})v^{(j)}(r_{12})f^{(k)}(r_{12}) \ ,
\label{pot:dCd} \\
\Delta C_{ev} &=
\bar{U}_d \dfrac{\rho_0}{\nu} \int d^3r_{12} l_v^2(r_{12})
 \nonumber \\
&\times
\sum_{i,j,j\prime ,k,l} K^{ijj\prime }K^{j\prime kl}A^lP_l
(4-2\delta_{j\prime 1}-\delta_{j1}-\delta_{l1})
f^{(i)}(r_{12})v^{(j)}(r_{12})f^{(k)}(r_{12}) \ ,
\label{pot:dCev} \\
\Delta C_{eu} &=
-\bar{U}_d \dfrac{\rho_0}{\nu} \int d^3r_{12}l_u^2(r_{12})
 \nonumber \\
&\times
\sum_{i,j,j\prime ,k,l} K^{ijj\prime }K^{j\prime kl}A^lB_l
(4-2\delta_{j\prime 1}-\delta_{j1}-\delta_{l1})
f^{(i)}(r_{12})v^{(j)}(r_{12})f^{(k)}(r_{12}) \ .
\label{pot:dCeu}
\end{align}
where
\begin{align}
\bar{U}_d &= \dfrac{\rho_0}{3} \int d^3 r_{13} \sum_{i=2,3}
A^i f^{(i)}(r_{13}) f^{(i)}(r_{13}) \ ,
\label{bUd} \\
\bar{U}_{e2}^v &= -\dfrac{\rho_0}{\nu} \int d^3r_{13} l_v^2(r_{13})
\left(
f^{(1)}(r_{13})f^{(2)}(r_{13})-
f^{(1)}(r_{13})f^{(1)}(r_{13})+
4 f^{(3)}(r_{13})f^{(3)}(r_{13}) \right) \ ,
\label{bUe2v} \\
\bar{U}_{e2}^u &= -\dfrac{3\rho_0}{\nu} \int d^3r_{13} l_u^2(r_{13})
f^{(2)}(r_{13})f^{(2)}(r_{13}) \ .
\label{bUe2u}
\end{align}
These expressions sum the commutator corrections which are linear in the vertex corrections $D1$ and $D2$.
The much smaller higher order terms have been disregarded.

\subsection{Kinetic energy}
\label{sub:KinEn}

We will adopt the Jackson--Feenberg (JF) identity to evaluate the kinetic energy~\cite{WFF}.
The advantage of using this form lies in the fact that the JF kinetic energy operator is mainly constructed by the sum of one-- and two--body operators,
the three--body operators being almost negligable.
Other forms, like the Pandharipande--Bethe or the Clark--Westhaus ones, have large three--body pieces,
and need to go beyond our two--body plus vertex corrections approximation.

The kinetic energy expectation value per particle is given by the sum of a one-- and
a two-body term:
\begin{equation}
\dfrac{\langle\hat{T}\rangle_2}{\langle\hat{N}\rangle_1} = T_1 + T_2 \ ,
\end{equation}
where $T_1$ gives the \emph{uncorrelated} BCS kinetic energy per particle,
\begin{equation}
T_1 = \dfrac{\hbar^2}{2m} \dfrac{\nu}{\rho}
 \int \dfrac{d^3k}{(2\pi)^3} k^2 v^2(k) \ .
\end{equation}

The JF $T_2$ energy is given by:
\begin{align}
T_2 &=
-\dfrac{\hbar^2}{4m} \rho \int d^3r_{12} \left[T_d^{JF}(r_{12})
-\dfrac{1}{\nu} \left(\dfrac{c_e}{c_d}\right)^2
 \left[T_{ev}^{JF}(r_{12})l_v^2(r_{12})
-T_{eu}^{JF}(r_{12})l_u^2(r_{12}) \right.\right.
 \nonumber \\
&-\left.\dfrac{1}{2}\left(T_{2v}^{JF}(r_{12})\nabla^2 l^2_v(r_{12})
 -T_{2u}^{JF}(r_{12})\nabla^2 l^2_u(r_{12})\right) ]\right]
 +\Delta T_d + \Delta T_{ev} + \Delta T_{eu} \ .
\label{eq:JF}
\end{align}
$T_d$ and $T_{ev}$, corresponding to the direct and exchange terms of
the normal case, are:
\begin{align}
T_d^{JF}(r_{12}) &= \sum_{i} A^i
\left(f^{(i)}(r_{12})\nabla ^2 f^{(i)}(r_{12})
-(\vec\nabla f^{(i)}(r_{12}))^2 \right) \ ,
 \nonumber \\
T_{ev}^{JF}(r_{12}) &= \sum_{i,k,l} K^{ikl}A^lP_l
\left(f^{(i)}(r_{12})\nabla^2 f^{(k)}(r_{12})
-\vec\nabla f^{(i)}(r_{12})\cdot\vec\nabla f^{(k)}(r_{12}) \right) \ ,
 \nonumber \\
T_{2v}^{JF}(r_{12}) &= \sum_{i,k,l} K^{ikl}A^lP_l
\left(f^{(i)}(r_{12})f^{(k)}(r_{12}) - \delta_{i1}\delta_{k1}\right) \ .
\label{eq:TJFnormal}
\end{align}
Similarly, the $u$--terms are:
\begin{align}
T_{eu}^{JF}(r_{12}) &= \sum_{i,k,l} K^{ikl}A^lB_l
\left(f^{(i)}(r_{12})\nabla^2 f^{(k)}(r_{12})
-\vec\nabla f^{(i)}(r_{12})\cdot\vec\nabla f^{(k)}(r_{12}) \right) \ ,
 \nonumber \\
T_{2u}^{JF}(r_{12}) &= \sum_{i,k,l} K^{ikl}A^lB_l
\left(f^{(i)}(r_{12})f^{(k)}(r_{12}) - \delta_{i1}\delta_{k1}\right) \ .
\label{eq:TJFU}
\end{align}

The commutators terms are calculated as for the potential:
\begin{align}
\Delta T_d &=
\dfrac{\hbar^2}{2m}(\bar{U}_d-\bar{U}_{e2}^v + \bar{U}_{e2}^u) \rho_0
\int d^3r_{12}\sum_{i=2,3}A^i \left(f^{(i)}(r_{12})\nabla ^2 f^{(i)}(r_{12})
- (\vec\nabla f^{(i)}(r_{12}))^2 \right) \ ,
\label{kin:dTd:JF} \\
\Delta T_{ev} &= -\dfrac{\hbar^2}{2m}
\bar{U}_d \dfrac{\rho_0}{\nu} \int d^3r_{12}\sum_{i,k,l} K^{ikl}A^lP_l
(3-\delta_{i1}-\delta_{k1}-\delta_{l1})
\label{kin:dTev:JF} \\
&\times \left(\left(f^{(i)}(r_{12})\nabla^2 f^{(k)}(r_{12})-\vec\nabla
f^{(i)}(r_{12})
\cdot\vec\nabla f^{(k)}_k(r_{12})\right)l_v^2(r_{12})
-\dfrac{1}{2}f^{(i)}(r_{12})f^{(k)}(r_{12})\nabla^2 l_v^2(r_{12})\right) \ ,
\nonumber \\
\Delta T_{eu} &= \dfrac{\hbar^2}{2m}
\bar{U}_d \dfrac{\rho_0}{\nu} \int d^3r_{12}\sum_{i,k,l} K^{ikl}A^lB_l
(3-\delta_{i1}-\delta_{k1}-\delta_{l1})
\label{kin:dTeu:JF} \\
&\times \left(\left(f^{(i)}(r_{12})\nabla^2 f^{(k)}(r_{12})-\vec\nabla
f^{(i)}(r_{12})
\cdot\vec\nabla f^{(k)}(r_{12})\right)l_u^2(r_{12})
-\dfrac{1}{2}f^{(i)}_i(r_{12})f^{(i)}_k(r_{12})\nabla^2
l_u^2(r_{12})\right) .\nonumber
\end{align}

We disregard the small three--body contributions to the JF kinetic energy.
As for the potential energy, the commutator terms include only cluster diagrams which are linear in the vertex corrections.

The energy expectation value, at the two--body order of the cluster
expansion, is:
\begin{equation}
E_2 = \dfrac{\langle\hat{H}\rangle_2}{\langle\hat{N}\rangle_1}
    = T_1 + T_2 + V_2 \ .
\label{eq:e2}
\end{equation}

\section{Euler and correlated gap equations}
\label{sec:Euler}

The Euler and the correlated gap equations form a set of coupled
equations, whose solution determines the correlation functions and the
correlated BCS amplitudes. They result from the variational requirement:
\begin{equation}
\delta_{v, f^{(i)}}
 \langle \hat{H} - \lambda \hat{N} \rangle \ = \ 0 \ .
\end{equation}

In deriving the equations we will use the two--body approximation
previously
discussed,
\begin{gather}
\langle \hat{H} \rangle \ \sim \ \langle \hat{H} \rangle_2 \ ,
  \nonumber \\
\langle \hat{N} \rangle \ \sim \ \langle \hat{N} \rangle_1 \ = \
\Omega\rho \ ,
\end{gather}
where $E_2$ and $\rho_1$ are given in Eqs. (\ref{eq:e2}) and
(\ref{eq:rho1}),
respectively.

We will make further approximations, which we believe are
accurate enough, but that can be eventually released. They consist of:
\begin{itemize}
\item[(i)]
neglecting the commutator terms in the derivation of the Euler equation,
while keeping them in the calculation of $E_2$ and $\rho_1$;
\item[(ii)]
decoupling the BCS amplitude, $v(q)$, from the correlation functions,
$f^{(i)}(r)$. As a consequence, we neglect the implicit dependence
on $f^{(i)}(r)$ in the functional variation with respect to $v(q)$, and
viceversa in the derivation of the Euler equations for $f^{(i)}(r)$.
\end{itemize}
In this way we arrive at an Euler equation of \emph{precisely} 
the same algebraic structure as that of the bare BCS scheme, 
 with the Hamiltonian containing paired
terms only \cite{Leggett:1975}. However, 
with respect to the ordinary BCS treatment,
there is the crucial distinction that the pairing force and
the single-particle energies are now \emph{renormalized} by 
the dynamical correlations.
Correlations also affect the mean density through the vertex corrections of the BCS/FHNC theory.
The explicit formulae are given below.

\subsection{Euler equations for the correlation functions}
\label{sub:Eul_CF}

Following the PW notation for standard nuclear matter, where the
Schr\"odinger--like equations are written in the $T,S$ channels (here
we consider the isospin $T = 1$ channel only), the following
changes are made with respect to the normal phase equations:

\begin{itemize}

\item[($S=0$) :\ ] in the singlet channel, eq.~(3.12) of PW,
\begin{equation}
\Phi_{S=0, T=1} \rightarrow \sqrt{1 +
 \left(\dfrac{c_e}{c_d}\right)^2 \left( l_v^2(r) + 2 l_u^2(r) \right) } \ .
\end{equation}

\item[($S=1$) :\ ] in the triplet channel, eq.~(3.14) of PW,
\begin{equation}
\Phi_{S=1, T=1} \rightarrow \sqrt{1 -
 \left(\dfrac{c_e}{c_d}\right)^2 l_v^2(r) } \ .
\end{equation}

\end{itemize}
The modifications to the spin--orbit equations are not included 
since we are dealing with a $v_6$ model.

\subsection{Correlated gap equation}
\label{sub:Eul_vq}

The correlated gap equations is derived from: 
\begin{equation}
\delta_{v(k)} (\rho E_2 - \rho\lambda) \ = \ 0 \ .
\end{equation}

The functional variation of the density is given by:
\begin{equation}
\delta_{v(k)} \rho \ = \ (1 + U_d + U_e) \delta_{v(k)} \rho_0 +
\rho_0 \delta_{v(k)} (U_d + U_e) \ ,
\label{eq:Drho}
\end{equation}
where
\begin{equation}
\delta_{v(k)} \rho_0 = \rho_0 \dfrac{\nu k^2}{2\pi^2\rho_0}
 \left(2v(k) \delta v_k \right) \ .
\label{eq:Drho0}
\end{equation}

After performing the tedious variations of the vertex terms, $U_d$ and
$U_e$,
we arrive at the expression:
\begin{equation}
\delta_{v(k)}\rho = \tilde E_0(k) \left(\delta_{v(k)}\rho_0\right) \ ,
\end{equation}
where
\begin{align}
\tilde E_0(k) \ =& \ \dfrac{\delta_{v(k)} \rho}{\delta_{v(k)} \rho_0}
 = 1 + U_d(3 - 4v^2(k)) + U_{e3}
\nonumber \\
 \ +& \ \dfrac{2 - v^2(k) - 4v^4(k)}{2u(k)v(k)} \int \dfrac{d^3q}{(2\pi)^3}
 F_u(q) u(|{\bm k} - {\bm q}|) v(|{\bm k} - {\bm q}|)
\nonumber \\
 \ +& \ \dfrac{1 - 2v^2(k)}{2u(k)v(k)} \int \dfrac{d^3q}{(2\pi)^3}
 F_u(q) u(|{\bm k} - {\bm q}|) v^3(|{\bm k} - {\bm q}|) \ ,
\end{align}
and $F_u(q)$ is
\begin{equation}
 F_u(q) = \int d^3r \exp (\imath {\bm q} \cdot {\bm r})
 \sum_{i,k,l} K^{ikl}A^l B_l
 \left(f_i(r)f_k(r)-\delta_{i1}\delta_{k1}\right) \ .
\end{equation}

The variation of $\rho$ plays a role in:

\begin{itemize}

\item[(i)]
the term $\rho\lambda$ of Eq. (\ref{eq:Drho}), giving rise to
$\lambda \tilde E_0(q) \left(\delta_{v(k)} \rho_0 \right)$;

\item[(ii)]
the direct term of $\ \rho E_2\rightarrow \rho E_d$, where
\begin{equation}
\rho E_d = \dfrac{\rho^2}{2} \int d^3r_{12}
 \left(V_d(r_{12}) - \dfrac{\hbar^2}{2m} T_d(r_{12}) \right) \ .
\end{equation}
In this case we get:
$2E_d \tilde{E}_0 \left(\delta_{v(k)} \rho_0\right)$;

\item[(iii)]
the exchange term of $\rho E_2 \rightarrow \rho E_e$, where
\begin{align}
\rho E_e =& -\dfrac{1}{2\nu} (c_e\rho_0)^2 \int d^3r_{12} \Bigl\{
 \left(V_{ev}(r_{12}) l_v^2(r_{12}) - V_{eu}(r_{12}) l_u^2(r_{12})\right)
\nonumber \\
 -& \dfrac{\hbar^2}{2m} \Bigl[
 T_{ev}(r_{12}) l_v^2(r_{12}) - T_{eu}(r_{12}) l_u^2(r_{12})
\nonumber \\
 -& \dfrac{1}{2} \left(T_{2v}(r_{12}) \nabla^2 l^2_v(r_{12}) -
 T_{2u}(r_{12})\nabla^2 l^2_u(r_{12})\right) \Bigr] \Bigr\} \ .
\end{align}
This variation applies to the $l_v(r)$ and
$l_u(r)$ functions appearing in $E_e$, with the result
\begin{equation}
\delta_{v(k)} (\rho E_e) = \left\{
 \Sigma(k) - \Delta(k) \dfrac{1 - 2v^2(k)}{2u(k)v(k)} \right\}
 \left( \delta_{v(k)} \rho_0 \right) \ ,
\end{equation}
where
\begin{align}
\Sigma(k) \ =& \ -c_e^2 \int \dfrac{d^3q}{(2\pi)^3}
 v^2(|{\bm k} - {\bm q}|)
 \left(V_{ev}^T(q) - \dfrac{\hbar^2}{2m} \left(T_{ev}^T(q) +
 \dfrac{q^2}{2} T_{2v}^T(q) \right)\right)
\nonumber \\
\ +& \ 2\dfrac{c_e - 1}{c_e}(c_dE_e) \ ,
\label{eq:Sigma} \\
\Delta(k) \ =& \ -c_e^2 \int \dfrac{d^3q}{(2\pi)^3}
 u(|{\bm k} - {\bm q}|)v(|{\bm k} - {\bm q}|)
 \left(V_{eu}^T(q) - \dfrac{\hbar^2}{2m}\left(T_{eu}^T(q) +
 \dfrac{q^2}{2} T_{2u}^T(q) \right)\right) \ ,
\label{eq:Delta}
\end{align}
where $V_{ev}^T(q)$ and $V_{eu}^T(q)$ are the Fourier transforms of $V_{ev}(r)$ and $V_{eu}(r)$
of Eqs.~(\ref{eq:vnormal}) and (\ref{eq:ucomponent}), respectively.
Similarly,
$T_{ev}^T(q)$ and $T_{2v}^T(q)$ are the Fourier transforms of $T_{ev}^{JF}(r)$ and $T_{2v}^{JF}(r)$ of Eq.~(\ref{eq:TJFnormal});
$T_{eu}^T(q)$ and $T_{2u}^T(q)$ are the Fourier transforms of $T_{eu}^{JF}(r)$ and $T_{2u}^{JF}(r)$ of Eq.~(\ref{eq:TJFU}).
\end{itemize}

Notice that $\Sigma(k)$, of Eq.~(\ref{eq:Sigma}), includes the constant
term provided by the functional variation of the vertex correction.

After collecting all the terms, we may write a correlated gap equation,
or Euler equation for the correlated BCS amplitudes, in
the form:
\begin{equation}
 \left(\dfrac{\hbar^2 k^2}{2m} - \bar{\lambda} \tilde E_0(k) \right) v(k)
 + \Sigma(k) v(k)
 - \Delta(k) \dfrac{1 - 2v^2(k)}{2\sqrt{1 - v^2(k)}} \ = \ 0 \ ,
\label{eq:bcseq}
\end{equation}
which resembles that obtained in standard BCS theory~\cite{Leggett:1975}.
The solution for $v^2(k)$ of the correlated gap equation can be written as:
\begin{equation}
v^2(k) = \dfrac{1}{2} \left(1 - \dfrac{\epsilon(k)}{E(k)} \right) \ ,
\label{eq:bcssol}
\end{equation}
with
\begin{gather}
\epsilon(k) \ = \
 \dfrac{\hbar^2 k^2}{2m} + \Sigma(k) -\bar{\lambda} \tilde E_0(k) \ ,
\label{eq:epsilon} \\
E(k) \ = \ \sqrt{\Delta^2(k) + \epsilon^2(k)} \ ,
\label{eq:E} \\
\bar{\lambda} \ = \ \lambda - 2E_d \ ,
\label{eq:lambda}
\end{gather}
where $\Delta(k)$ has to be interpreted as the \emph{correlated} gap
function.  Its value at $k=k_{\text F}$, $\Delta_{\text F}$, 
is the energy gap, namely the energy 
required to break a pair at the Fermi surface. 
In the present case, the functions $\tilde E_0(k)$, $\Sigma(k)$
and $\Delta(k)$ all depend upon $v^2(k)$, making the correlated gap equation
(\ref{eq:bcssol}) highly non linear.

\subsection{Correlated versus standard BCS equation}
\label{sub:BCStand}

The correlated BCS equation~(\ref{eq:bcssol}) has the same algebraic
structure
as the uncorrelated one (see ref.~\cite{Leggett:1975}, Eqs.~(5.29) and
(5.30)).
However, the standard BCS equations do not contain the $\Sigma(k)$ term,
whereas in our
approach $\Sigma(k)\neq 0$, even if the correlation operator is set equal
to $1$.
In fact, in this case the quantities $\tilde E_0(k)$, $\Sigma(k)$ and
$\Delta(k)$ become:
\begin{align}
\tilde E_0(k) \ \rightarrow& \ 1 \ ,
\nonumber \\
\Sigma(k) \ \rightarrow& \
 - \int \dfrac{d^3q}{(2\pi)^3} v^2(|{\bm k} - {\bm q}|)
 \left[v_c(q) + 3v_\sigma(q) \right] \ ,
\nonumber \\
\Delta(k) \ \rightarrow& \
 - \int \dfrac{d^3q}{(2\pi)^3}
 u(|{\bm k} - {\bm q}|) v(|{\bm k} - {\bm q}|)
 \left[v_c(q) - 3v_\sigma(q)\right] \ .
\label{BCS0}
\end{align}


In correlated BCS, $\Sigma(k)$ \emph{dresses} the single particle energies
$\hbar^2k^2/2m$, and $\tilde E_0(k)$ \emph{renormalizes} the mass.

Similarly, $\Delta(k)$ assumes the role of the gap function.
From Eq.~(5.32) of ref.~\cite{Leggett:1975}
\begin{equation}
 \Delta(k) =
 -\sum V_{{\bm k} {\bm q}} \dfrac{\Delta(q)}{2E(q)} \ ,
\end{equation}
where
\begin{equation}
 V_{{\bm k} {\bm q}} =
 \int d^3 r \ \exp({\imath({\bm k} - {\bm q}) \cdot {\bm r}}) V(r) \, .
\label{eq:FTF}
\end{equation}
From Eq.~(\ref{eq:bcssol}) and the normalization relation,
$u^2(k) + v^2(k) = 1$, it follows that
\begin{equation}
u(k)v(k) = \dfrac{1}{2} \sqrt{1 - \dfrac{\epsilon^2(k)}{E^2(k)}}
 = \dfrac{\Delta(k)}{2E(k)} \ .
\end{equation}
Therefore
\begin{equation}
\Delta(k) =
 -\sum V_{{\bm k} {\bm q}} u(q) v(q) \ ,
\end{equation}
coincides with Eq.~(\ref{eq:Delta}).

The comparison with the uncorrelated BCS theory allows
identifying $E(k)$ with the excitation energy of the
broken pair (BP) with respect to the ground--state,
as defined in Ref.~\cite{Leggett:1975},
\begin{equation}
 E_\text{BP} - E_\text{GS} \ \equiv \ E(k) \ .
\end{equation}

\section{Results}
\label{results}

We have solved the BCS and correlated BCS equations for neutron
matter with a variety of potentials, namely the Reid (R),
Argonne $v_{14}$ (A14) and Argonne $v_{8^\prime}$ (A8$^\prime$) ones.
In solving the gap equations we have generalized the method described
by Khodel et al. in Ref.~\cite{Khodel:1996}.
According to this method the original gap equation is identically
replaced by a
set of coupled equations: a non-singular quasilinear integral equation
for the dimensionless profile function, $\chi(k)$, defined by
$\Delta(k) = \chi(k)\,\Delta_\text{F}$ and a non-linear algebraic one
for the gap, $\Delta_\text{F} = \Delta(k_\text{F})$, at the Fermi surface.
After integrating Eq.~(\ref{eq:Delta}) over the angle , we obtain
\begin{equation}
 \Delta(k) \ = \ -c_e^2 \int_0^\infty \dfrac{q^2dq}{2\pi^2}
 \dfrac{V(k,q) \, \Delta(q)}
 {2\sqrt{\Delta^2(q) + \epsilon^2(q)}} \, ,
\label{BCS-k}
\end{equation}
with,
\begin{align}
 V(k,q) \, =& \,
 4\pi\int_0^\infty r^2dr j_0(kr)\left\{V_{eu}(r) - \dfrac{\hbar^2}{2m}
 \left[T_{eu}(r) + \dfrac{k^2+q^2}{2} T_{2u}(r)\right]\right\} j_0(qr)
\nonumber \\
  +& \,
 \dfrac{\hbar^2 kq}{2m} 4\pi\int_0^\infty r^2dr j_1(kr) T_{2u}(r)
j_1(qr) \, .
\label{BCS-V}
\end{align}
It is assumed that the interaction $V(k,q)$ is different from zero at
the Fermi surface, $V(k_\text{F},k_\text{F}) \ne 0$. To solve the gap
equation we decompose the potential, $V(k,q)$, into a separable part
and a remainder, $W(k,q)$, that vanishes when either argument is at
the Fermi surface:
\begin{equation}
 V(k,q) \ = \ V_\text{F} \phi(k) \phi(q) + W(k,q) \ ,
\label{PotDec}
\end{equation}
where $W(k_\text{F}, k) = W(k, k_\text{F}) \equiv 0$ and
$\phi(k) = V(k, k_\text{F}) / V_\text{F}$. Then, the gap equation
(\ref{BCS-k})
is readily seen to be equivalent to an integral equation for the shape
function, $\chi(k)$,
\begin{equation}
 \chi(k) + c_e^2 \int_0^\infty \dfrac{q^2dq}{2\pi^2}
 \dfrac{W(k, q) \, \chi(q)}
 {2\sqrt{\Delta_\text{F}^2 \chi^2(q) + \epsilon^2(q)}} = \phi(k) \ ,
\label{BCS:shape}
\end{equation}
together with the {\em algebraic} equation,
\begin{equation}
 1 + c_e^2 V_\text{F} \int_0^\infty \dfrac{q^2dq}{2\pi^2}
 \dfrac{\chi(q) \, \phi(q)}
 {2\sqrt{\Delta_\text{F}^2 \chi^2(q) + \epsilon^2(q)}} = 0 \ ,
\label{BCS:D_F}
\end{equation}
for the gap amplitude $\Delta_\text{F}$ (assumed nonzero).
Since $W(k, k_\text{F})$ is zero by construction, the integral equation
(\ref{BCS:shape}) has a nonsingular kernel, the log-singularity of the BCS
equation having been isolated in the amplitude equation (\ref{BCS:D_F}).
An iterative solution of this set of equations converges very rapidly.

The correlated gap equations are solved using the BCS
solution at a given $k_\text{F}$ as an input. We find that the final
density, $\rho$, is always very close to the initial one, $\rho_0$.
The maximum difference between $k_\text{F}^{input}$ and
$k_\text{F}^{final}$ is well below one percent.
In Table (\ref{Tab:den}) we show the input and output values of
$k_\text{F}$, of the density, $\rho$, of the chemical potentials,
$\lambda_\text{F}$, of the effective mass, $m^\star/m$,
defined by the relation:
\begin{equation}
 \left( \dfrac{m^\star}{m} \right)^{-1} \, = \,
 \dfrac{m}{\hbar^2} \left( \dfrac{1}{k} \dfrac{d e(k)}{d k}
\right)_{k=k_\text{F}} \,
 \, , \, \, e(k)\, = \, \dfrac{\hbar^2 k^2}{2m} + \Sigma(k) \, ,
\end{equation}
and of the gap $\Delta_F$ obtained with the A8$^\prime$ model for the
uncorrelated BCS, and for the Jastrow (J) and $f_6$ correlated (CO) cases.

It is evident that the introduction of the correlations very slightly
affects the total density. On the contrary the chemical potential is
reduced by the Jastrow correlations by $\sim 20$ to $\sim 30\%$.
Spin dependent correlations provide a further, even if small, decrease
of $\lambda_F=\hbar^2 k_\text{F}^2/2m$. The effective mass, computed via the self--energy
$\Sigma (k)$, considerably decreases after the introduction of the
correlations. The normal phase effective mass,
computed microscopically in CBF, at $k_F = 0.8~\text{fm}^{-1}$ is
$\sim 0.8$. 

\begin{figure}[ht]
\includegraphics[angle=0, scale=0.62]{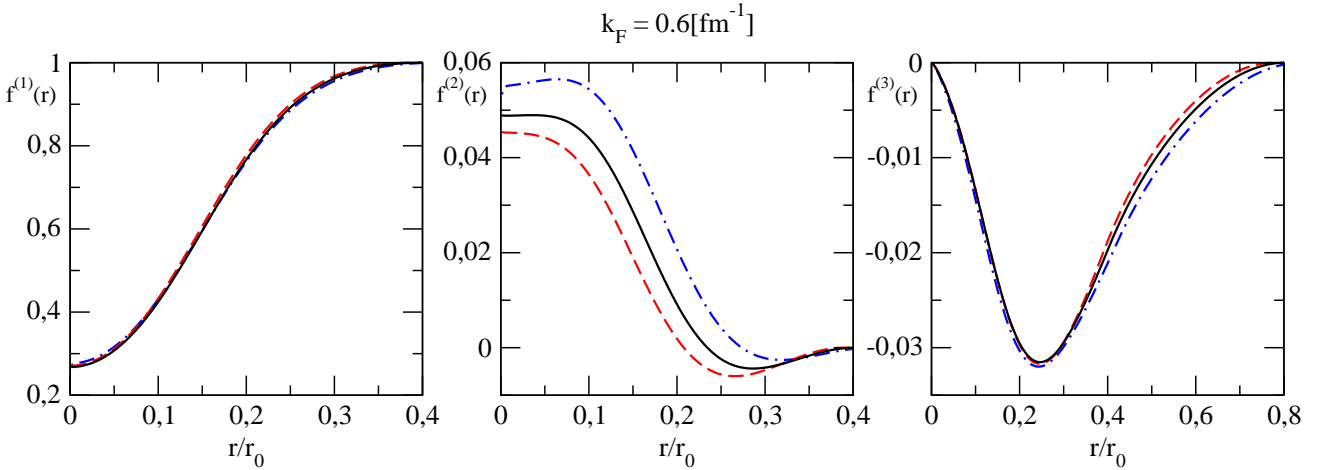}
 \caption{\label{Fig:Corr06}%
(colored online)
The central, spin and tensor correlation functions at the
$k_\text{F} = 0.6~\text{fm}^{-1}$ for the A8$^\prime$ potential.
The dash-dotted, blue lines are the normal phase correlations.
The dashed and solid, black lines are the BCS correlations without and with vertex corrections, correspondingly.
}%
\end{figure}

 In Figure~(\ref{Fig:Corr06}) we show the Jastrow, spin and tensor
correlations at $k_\text{F} = 0.6~\text{fm}^{-1}$ for the A8$^\prime$
potential.
The dash-dotted lines are the normal phase correlations, whereas the
solid lines give the correlations after solving the correlated
gap equations. For the $S$--pairing case Jastrow and tensor correlations
do not change from the normal to the BCS phases. Instead, the spin
correlation shows some sensitivity to the environmental phase. It is
reasonable to expect that for the $^3P_2$--$^3F_2$ pairing the tensor
correlation also will depend on the phase.
 
\begin{figure}[ht]
\vspace{0.5cm}
\includegraphics[angle=0, scale=0.7]{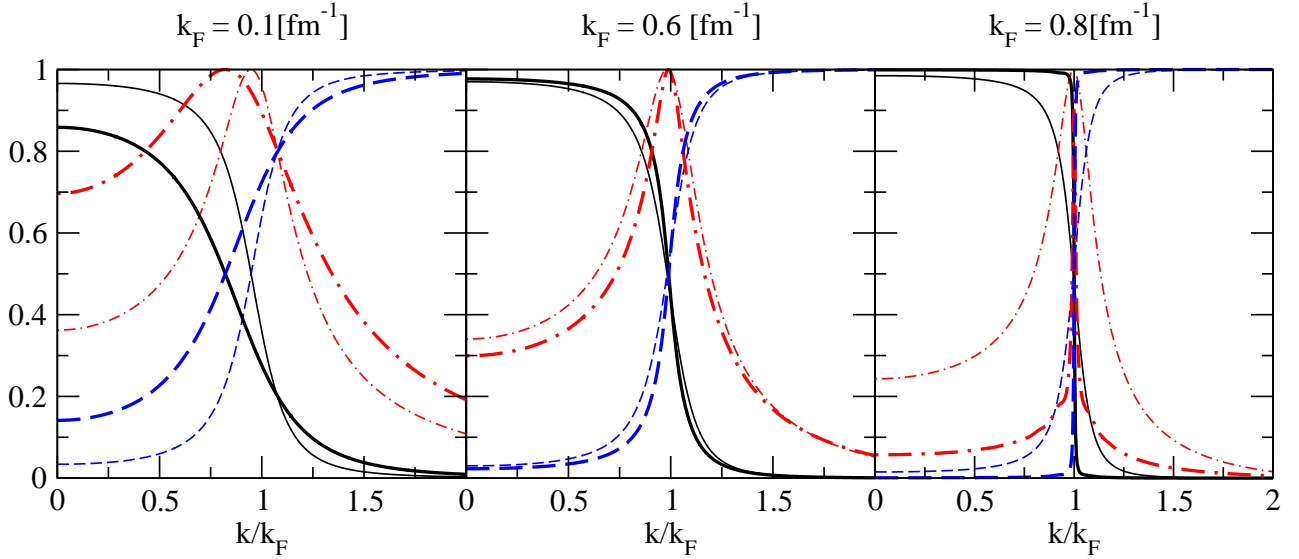}
 \caption{\label{Fig:v2_2vu_u2}%
(colored online)
$v^2(k)$ (solid, black lines), $2v(k)u(k)$ (dash-dotted, red lines) and
$u^2(k)$ (dashed, blue lines) amplitudes obtained from the
A8$^\prime$ potential at three densities.
The thin lines are the pure BCS results;  
the thick ones represent the correlated BCS amplitudes.
}%
\end{figure}

The $u^2$, $2uv$ and $v^2$ amplitudes, both for the pure and correlated
BCS cases,
are shown in Figure~(\ref{Fig:v2_2vu_u2}) at three Fermi momenta.
At the lowest value, $k_\text{F} = 0.1~\text{fm}^{-1}$, the uncorrelated
and
correlated amplitudes substantially differ among each other, the correlated
ones showing a larger deviation from the step function, consistent with
the larger gap value ($\Delta^0_\text{F} = 0.07~\text{MeV}$ and
$\Delta^\textrm{CO}_\text{F} = 0.14~\text{MeV}$). At $k_\text{F} =
0.6~\text{fm}^{-1}$
the amplitudes are very close in both approaches, yielding similar gaps
($\Delta^0_\text{F} = 2.27~\text{MeV}$ and
 $\Delta^\textrm{CO}_\text{F} = 2.25~\text{MeV}$). At the largest value,
$k_\text{F} = 0.8~\text{fm}^{-1}$, the correlated amplitudes are
practically step functions. 
In fact this is almost the highest density for which we
find solution to the correlated BCS equations.

\begin{figure}[ht]
\includegraphics[angle=0, scale=1.0]{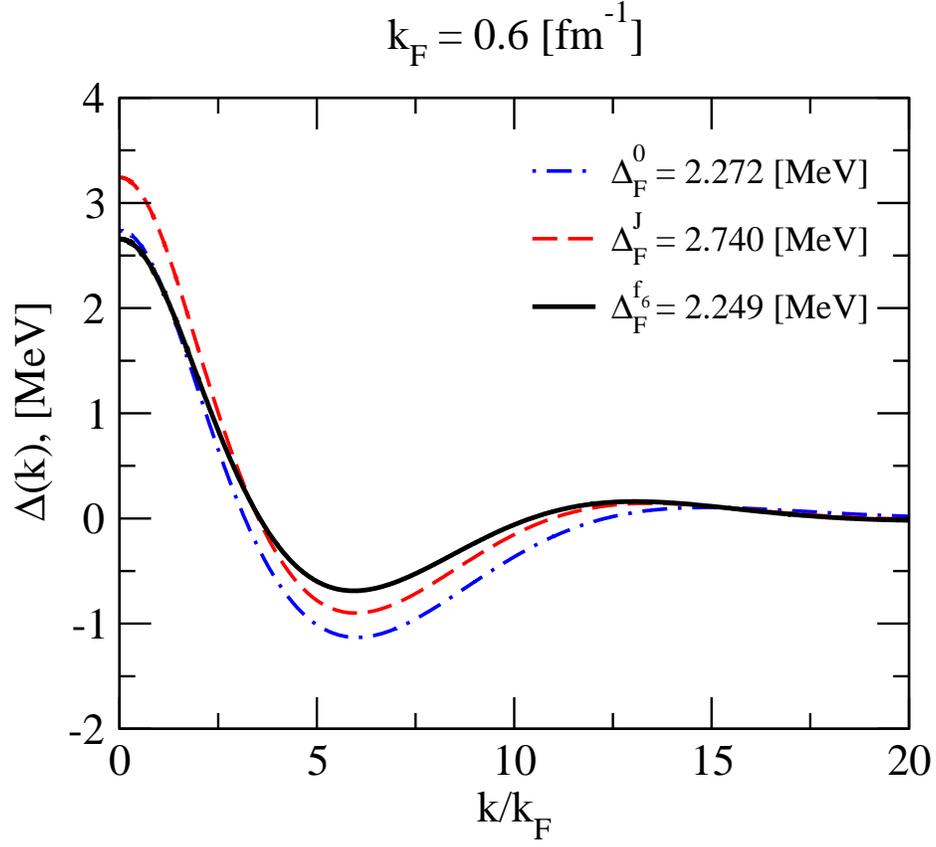}
\caption[]{\label{Fig:gapk-06}%
(colored online)
$\Delta(k)$, in MeV, obtained for the A8$^\prime$ model at
$k_\text{F} = 0.6~\text{fm}^{-1}$.
The dash-dotted, blue line is the pure BCS solution; 
the dashed, red line represents the Jastrow correlates BCS case; 
the solid, black line is the fully correlated BCS solution.
}%
\end{figure}

The gap function, $\Delta(k)$, at $k_\text{F} = 0.6~\text{fm}^{-1}$ is
given
in Figure~(\ref{Fig:gapk-06}). In addition to the pure and
correlated BCS functions, we show the one obtained by a simple
Jastrow--correlated wave--function. At low $k$--values, $k/k_\text{F} <
2.5$,
the effects of the Jastrow and spin--dependent correlations compensate,
providing a
gap function close to the BCS result. At larger momenta they add and
the correlated gap function departs from the uncorrelated one, up to
$k/k_\text{F} \sim 15$, where all functions have essentially vanished.

\begin{figure}[ht]
\vspace{1.cm}
\includegraphics[angle=0, scale=1.0]{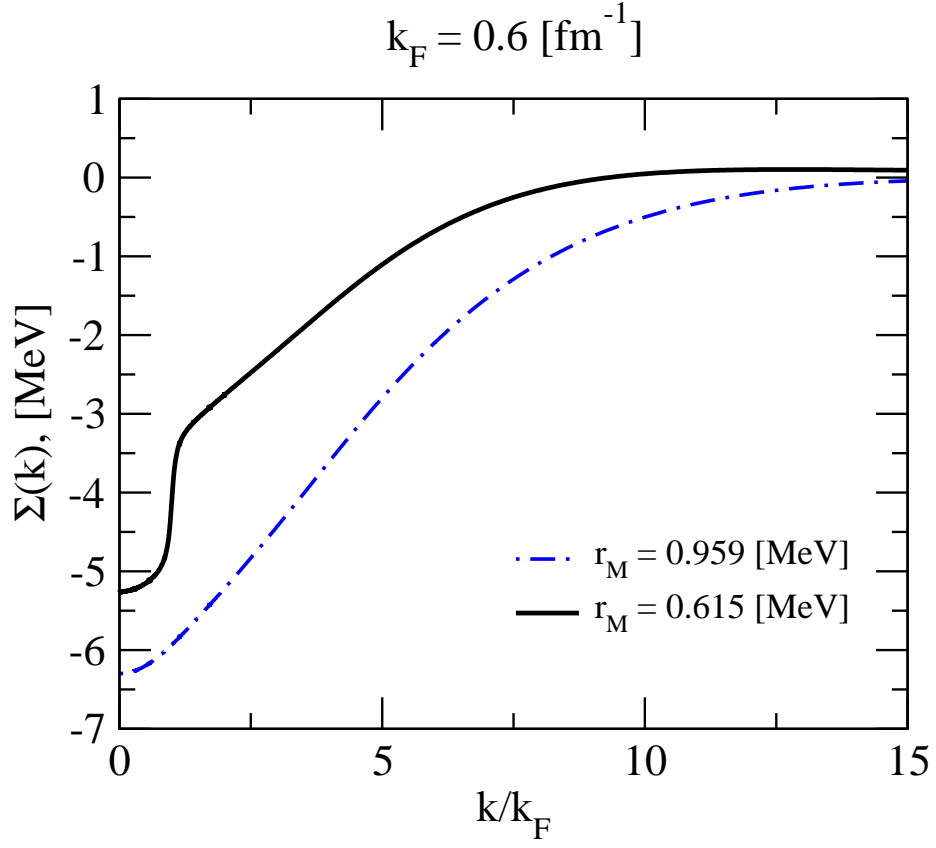}
\caption[]{\label{Fig:Sigma-06}%
(colored online)
The self-energy $\Sigma(k)$, in MeV, for the BCS (dash-dotted, blue
line) and correlated BCS (solid, black line) cases
for the A8$^\prime$ model at $k_\text{F} = 0.6~\text{fm}^{-1}$.
}%
\end{figure}

The self--energy, $\Sigma (k)$, is depicted in Figure~(\ref{Fig:Sigma-06})
at $k_\text{F} = 0.6~\text{fm}^{-1}$. The main difference between the
BCS and correlated BCS cases lies in the sharp rising of
the correlated $\Sigma (k)$ at $k\sim k_\text{F}$, which produces the
much lower effective mass given in Table~(\ref{Tab:den}),
$(m^\star/m)^0 = 0.96$ and $(m^\star/m)^\textrm{CO} = 0.62$.
The mass renormalization, caused by short-range correlations,
enhanced the dispersive effect of the mean field, which leads to
quenching of the energy gap, which is enhanced by the screening effect of
the neutron pairing potential.

\begin{figure}[ht]
\vspace{1.cm}
\includegraphics[angle=0, scale=0.9]{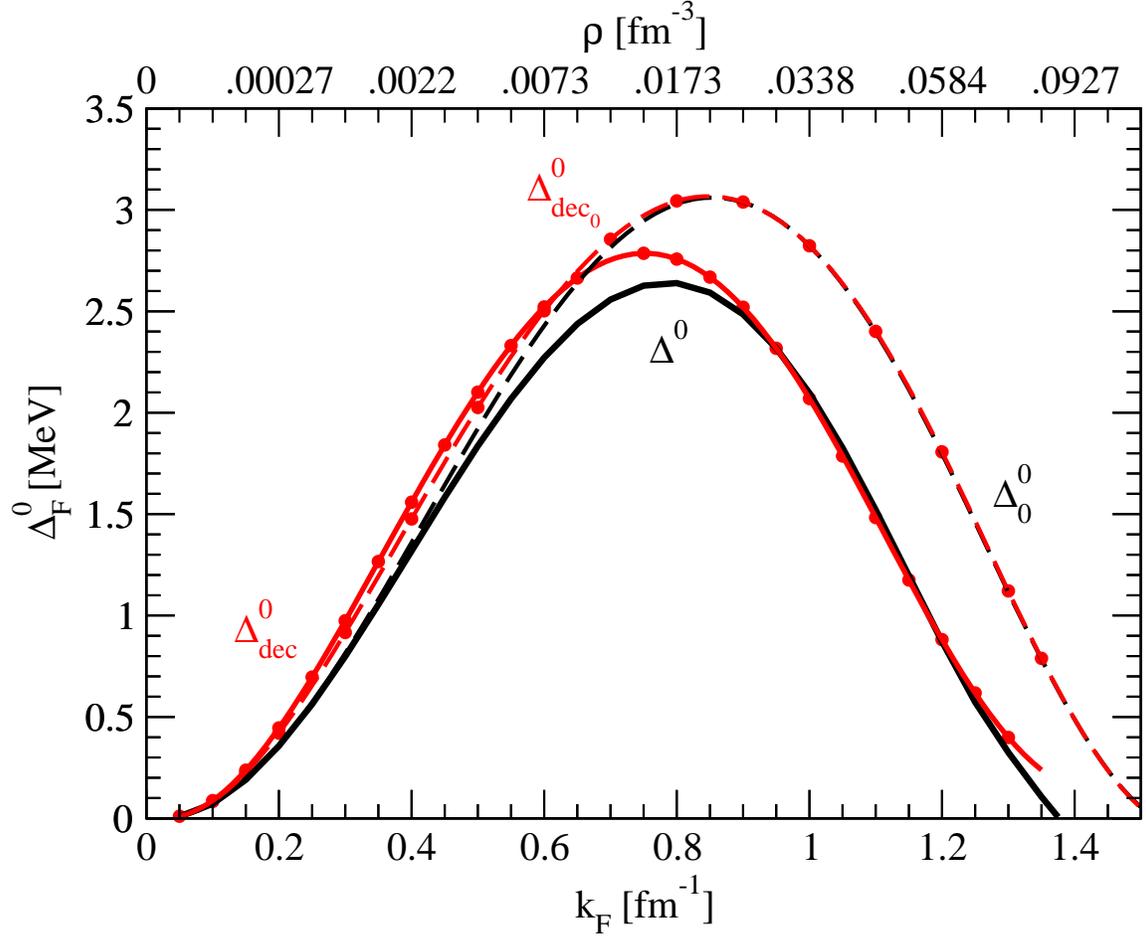}
\caption[]{\label{Fig:trial}%
$^1S_0$ pairing gaps obtained from the A8$^\prime$ potential in the
uncorrelated BCS case.
The curves labeled $\Delta^0_{\text{dec}_0}$ and $\Delta^0_\text{dec}$
 are calculated in the decoupling approximation, without and with
self--energy insertions, respectively. The remaining curves
refer to the full solution ($\Delta^0_0$ with $\Sigma(k)=0$ and
 $\Delta^0$ with $\Sigma(k)\ne 0$).
}%
\end{figure}

Figure~(\ref{Fig:trial}) displays the energy gap at the Fermi surface,
$\Delta_\text{F}^{(0)}$, as a function of the Fermi momentum for the
A8$^\prime$ potential in the uncorrelated BCS case. The curves correspond
to the full and to the decoupling approximation solutions of
Ref.~\cite{Decoupl,Chen:1986},
with and without the self--energy insertions of eq.~(\ref{BCS0}). 
The two gaps
are very close for $\Sigma (k) = 0$, whereas, after the introduction of
the self--energy, the decoupling approximation appears to slightly
overestimate the full solution.

\begin{figure}[ht]
\vspace{1.cm}
\includegraphics[angle=0, scale=0.9]{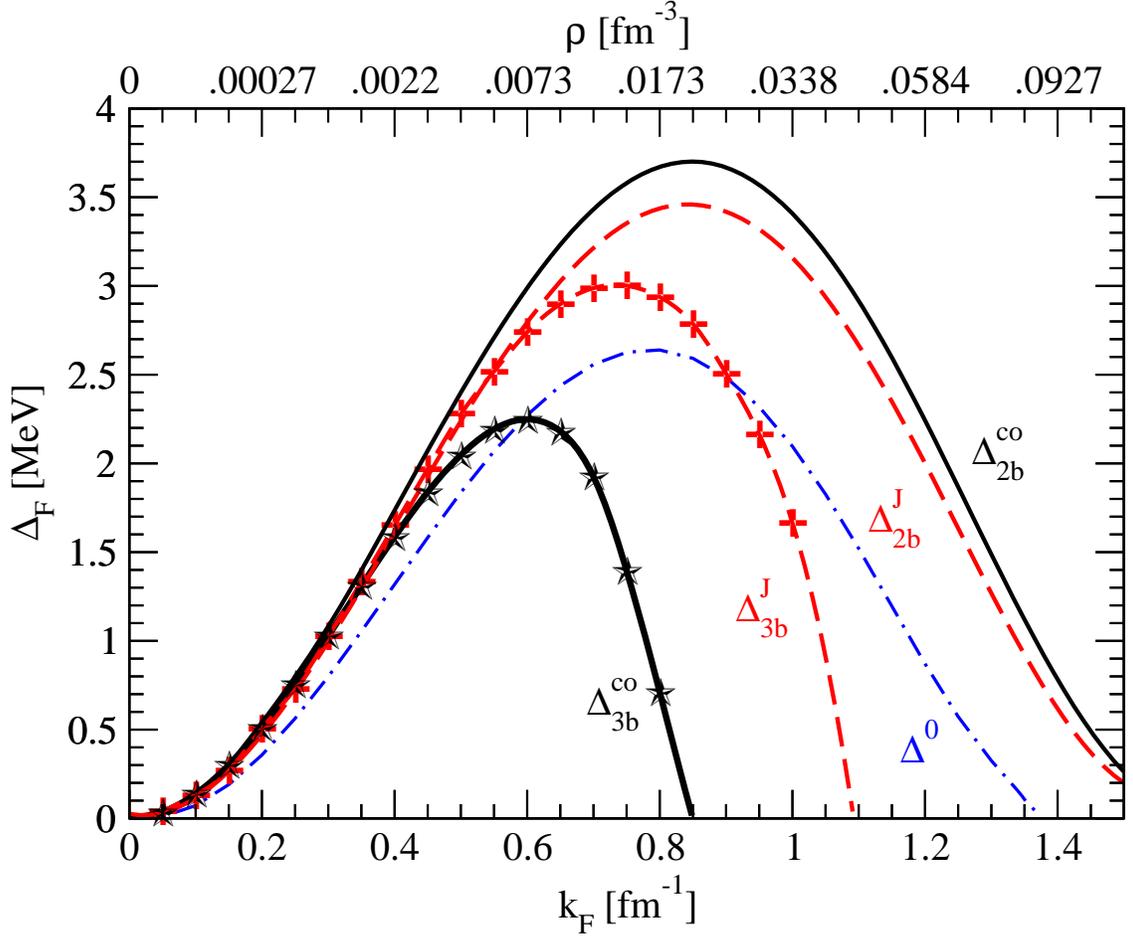}
\caption[]{\label{Fig:v12}%
A8$^\prime$ $^1S_0$ pairing gaps for different correlations and levels of
the cluster expansions. See text.
}%
\end{figure}

Figure~(\ref{Fig:v12}) gives the gaps for different types of correlations
(Jastrow and $f_6$) and at various levels of the cluster expansion for
the same potential.
The $\Delta^0$ gaps are the standard BCS results, those with the
superscript `J' are obtained within the Jastrow correlated theory and
the `\textrm{CO}' superscript denotes the corresponding correlations.
The $2b$ and $3b$ subscripts in the correlated gaps refer to the pure
two--body cluster case and to the one in which the density and the
vertex corrections are computed at the first order of the power
series expansion of Fig.~(\ref{Pic:BCS_diagrams}). The inclusion of the
Jastrow and $f_6$ correlations in the $2b$ case enhance the gap,
because the short--range repulsion of the potential
is renormalized by the short--range correlations. The $3b$ cases
include medium modification effects via higher order cluster
terms. Their effect is quite sizeable and reverse the behavior, 
both reducing the density region where
we find a BCS solution and decreasing the maximum gap with respect
to the standard case for the spin--dependent correlations. In fact,
$\Delta^0(\text{max}) \sim 2.6~\text{MeV}$ at $k_\text{F} \sim
0.9~\text{fm}^{-1}$,
while $\Delta^{3b}_{f_6}(\text{max}) \sim 2.2~\text{MeV}$ at
$k_\text{F} \sim 0.6~\text{fm}^{-1}$. These results indicate that 
higher order many--body cluster terms may be relevant to estimate 
the gap.

\begin{figure}[ht]
\vspace{1.cm}
\includegraphics[angle=0, scale=0.9]{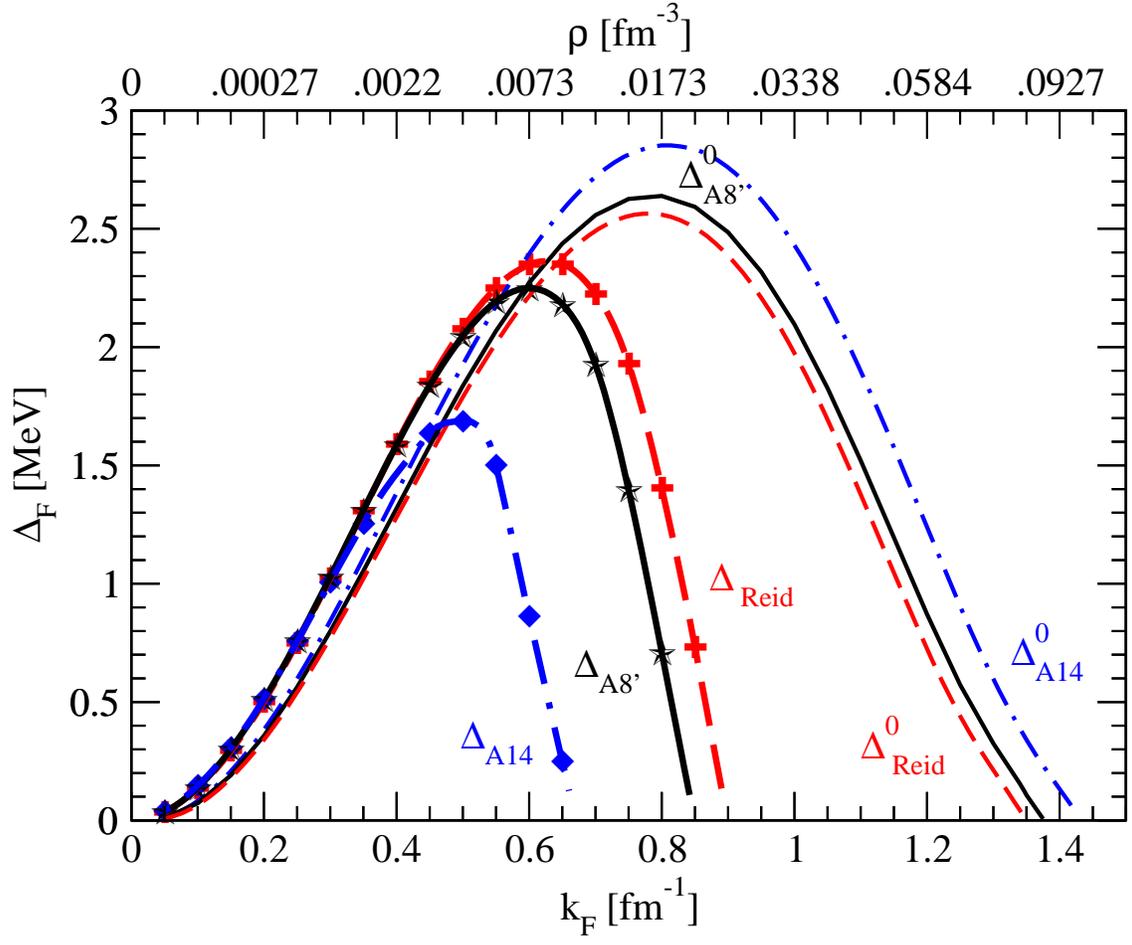}
\caption[]{\label{Fig:v14}%
$^1S_0$ pairing gaps for different nucleon-nucleon potentials.
}%
\end{figure}

Finally, in Figure~(\ref{Fig:v14}) we show the gaps for different
potentials in the BCS and $f_6$--correlated theories. We have used,
besides the Argonne $v_{8^\prime}$ model, also the Reid and Argonne
$v_{14}$
(A14) potentials. These potentials differ mostly for the strength of the
one--pion exchange induced components. In fact, A14 has much stronger
spin and tensor potentials than Reid and A8$^\prime$. This difference shows
up in the gaps, in both approaches. The BCS gap is larger in A14 than
for the other potentials, and more drastically reduced in the correlated
case, where $\Delta_{\text{A14}}(\text{max}) \sim 1.7~\text{MeV}$ at
$k_\text{F} \sim 0.5~\text{fm}^{-1}$.

Our results for the A14 potential are qualitatively similar to those of
Ref.~\cite{Lombardo:2001} (see also Ref.~\cite{Cao:2006}, where the more recent calculations was done),
where the medium polarization was included via Landau theory. The authors found an
analogous decrease of the BCS gap,
with $\Delta_{\text{A14}}^\textrm{Landau}(\text{max}) \sim 1.5-2~\text{MeV}$
at $k_\text{F}\sim 0.8~\text{fm}^{-1}$, but with a wider density region
allowing for a superfluid solution.


\section{Conclusions and perspectives}
\label{conclusions}

The problem of an accurate determination of the BCS gap in a strongly
interacting matter of nucleons is a longstanding one. Medium modification
effects are expected to be important, but of difficult quantitative
evaluation. We have used FHNC/BCS theory to take
care of the short range correlations induced by the interaction in
neutron matter at zero temperature. We have adopted the realistic Argonne
$v_{8^\prime}$
two--nucleon potential and a correlation factor having central, spin
and tensor dependent components. The density has been computed at the
first order of the power series expansion, since this expansion provides
at each order the correct density normalization in the normal
phase. Consistently, the matrix elements of the hamiltonian in the
correlated BCS state are evaluated at the two--body cluster level plus
vertex corrections at the interacting pair. This treatment, in conjunction
with the use of spin and tensor correlations lowers the maximum gap
at $k_\text{F}$ by $\sim 20\%$ with respect to the uncorrelated BCS case.
Moreover, $\Delta_\text{F}$ is shifted to a lower density. It is
clear from our results the relevance of state dependent correlations
for a reliable estimate of $\Delta_\text{F}$ in neutron matter,
as well as the need for inserting medium modifications via higher
order terms of the cluster expansion. Simple Jastrow, spin independent
correlations always enhance $\Delta_\text{F}$, even if massive summations
of cluster diagrams are performed. This effect is due to the screening of
the short--range repulsive interaction provided by the Jastrow
correlations.
Similar conclusions are drawn when the short--range correlations are
introduced by medium effects
within the Brueckner
G--matrix theory. State dependent correlations reverse this scenario and,
after the inclusion of the vertex corrections, reduce $\Delta_\text{F}$.
A qualitatively analogous result is found when state dependence is
introduced by a CBF based perturbative expansion theory on top of
Jastrow correlated states, but with spin dependent interactions.
 
In conclusion, we have stressed in this paper the importance of
state dependent correlations and medium effects
in superfluid neutron matter.
Both of these tend to reduce the $^1S_0$ pairing gap, confirming previous
studies.

 The large effects found either by extending the study to the full FHNC/SOC calculations or after the introduction of the vertex corrections, 
strongly point to the need of a realistic estimate of many body effects.  
This can be done by using the calculated correlated BCS amplitudes as the guiding function of an AFDMC calculation.
In the latter approach it is crucial to have a realistic guiding function in the path constraint.  
An extension of AFDMC to deal with the superfluid phases of neutron matter has been recently made
and preliminary results, obtained for $14$ neutrons, are given in Ref.~\cite{PRL_BCS}. 
A full description of both the AFDMC/BCS method and the corresponding results obtained for large systems will be given in a forthcoming paper~\cite{QMC_BCS}. 
An important issue is the role of the long range correlations.
This can be most easily done in a Jastrow correlated BCS case.
Work in this direction is in progress.


\section*{Acknowledgments}
This work has been partially supported by
the Italian MIUR through the PRIN:
\emph{Fisica Teorica del Nucleo Atomico e dei Sistemi a Molti Corpi}.
K.E.S. acknowledges partial support by the US National Science Foundation
via grant PHY-0456609.
A.Yu.I. is grateful to INFN and to the Dipartimento di Fisica ``E.Fermi''
of the University of Pisa and acknowledges partial support from the PRIN 2006
\emph{Quantum noise in mesoscopic systems}. 



\begin{table}
\begin{ruledtabular}
\begin{tabular}{lc|cccccccccccccc}
$k_\text{F}^0$ & $[\text{fm}^{-1}]$
 & 0.1 & 0.2 & 0.3 & 0.4 & 0.5 & 0.6 & 0.7
 & 0.8 & 0.9 & 1.0 & 1.1 & 1.2 & 1.3 \\
$\rho_0$ & $[\text{fm}^{-3}]$
 & .000034 & .00027 & .00091 & .00216 & .00422 & .00730 & .01158
 & .01729 & .02462 & .03377 & .04495 & .05836 & .07420 \\
$\lambda_\text{F}^0$ & $[\text{MeV}]$
 & .1780 & .6615 & 1.4771 & 2.6628 & 4.2509 & 6.2696 & 8.7478
 & 11.721 & 15.239 & 19.379 & 24.254 & 30.030 & 36.907 \\
$(m^\star/m)^0$ &
 & .9994 & .9960 & .9897 & .9811 & .9708 & .9589 & .9452
 & .9296 & .9120 & .8926 & .8716 & .8496 & .8270 \\
$\Delta_\text{F}^0$ & $[\text{MeV}]$
 & .0719 & .3568 & .7986 & 1.3188 & 1.8359 & 2.2719 & 2.5576
 & 2.6391 & 2.4850 & 2.0963 & 1.5209 & .8713 & .3247 \\
\hline
$k_\text{F}^\text{J}$ & $[\text{fm}^{-1}]$
 & .1001 & .2002 & .3006 & .4011 & .5020 & .6030 & .7041
 & .8049 & .9048 & 1.0032 & & & \\
$\rho_\text{J}$ & $[\text{fm}^{-3}]$
 & .000034 & .00027 & .00092 & .00218 & .00427 & .00741 & .01179
 & .01761 & .02502 & .03410 & & & \\
$\lambda_\text{F}^\text{J}$ & $[\text{MeV}]$
 & .1289 & .5100 & 1.1721 & 2.0775 & 3.2336 & 4.6656 & 6.3312
 & 8.2222 & 10.3590 & 12.737 & & & \\
$(m^\star/m)^\text{J}$ &
 & .9774 & .9480 & .9072 & .8609 & .8065 & .7370 & .6516
 & .5426 & .4035 & .2591 & & & \\
$\Delta_\text{F}^\text{J}$ & $[\text{MeV}]$
 & .1319 & .5067 & 1.0236 & 1.6511 & 2.2785 & 2.7398 & 2.9884
 & 2.9373 & 2.5069 & 1.6666 & & & \\
\hline
$k_\text{F}^{\text{CO}}$ & $[\text{fm}^{-1}]$
 & .1001 & .2002 & .3006 & .4011 & .5017 & .6022 & .7020
 & .8004 & & & & & \\
$\rho_{\text{CO}}$ & $[\text{fm}^{-3}]$
 & .000034 & .00027 & .00092 & .00218 & .00427 & .00738 & .01168
 & .01732 & & & & & \\
$\lambda_\text{F}^{\text{CO}}$ & $[\text{MeV}]$
 & .1204 & .4808 & 1.0709 & 1.8602 & 2.8082 & 3.8527 & 4.9220
 & 5.9532 & & & & & \\
$(m^\star/m)^{\text{CO}}$ &
 & .9787 & .9470 & .9023 & .8409 & .7533 & .6146 & .3997
 & .2476 & & & & & \\
$\Delta_\text{F}^{\text{CO}}$ & $[\text{MeV}]$
 & .1379 & .5104 & 1.0291 & 1.5861 & 2.0471 & 2.2487 & 1.9257
 & 0.7098 & & & & & \\
\end{tabular}
\end{ruledtabular}
\caption {
Fermi momentum, $k_\text{F}$, density, $\rho$, chemical potential,
$\lambda_\text{F}$, effective mas, $m^\star/m$, and gap value,
$\Delta_\text{F}$, in different approximations (see text).
}
\label{Tab:den}
\end{table}

\bibliographystyle{apsrev}
\bibliography{S-BCS}

\end{document}